# Nanozyme-based biosensing for clinical diagnosis of COVID-19: A mini review

Saeed Reza Hormozi Jangi

Hormozi Laboratory of Chemistry and Biochemistry, Zabol 9861334367, Iran; E-mail: saeedrezahormozi@gmail.com (S.R. Hormozi Jangi)

**Abstract**

Several clinical methods had been utilized for diagnosis of COVID-19 for instance, real-time reverse transcription-polymerase chain reaction (rRT-PCR), hematology examination, polymerase chain reaction (PCR), diagnostic guidelines based on clinical features, and Chest CT scans. However, the accurate current methods are time-consuming and expensive and other methods are inaccurate. To solve these drawbacks, nanozyme-based sensors have been developed for the reliable, accurate, and rapid detection of SARS-CoV-2. The main basis of these sensors is the detection of color variation of a nanozyme-mediated oxidation reaction in the presence and the absence of antigens of COVID-19. Besides, some of methods are based on probing the fluorescence of these systems as the clinical signal toward detection of SARS-CoV-2. This mini review focused on overviewing the nanozymes-based methods toward COVID-19 diagnosis. The historical background of COVID-19 was reviewed. Thereafter, the biomedical applications of nanozymes was discussed and finally, the recent progress of early diagnosis of COVID-19 based on nanozymatic systems was briefly reviewed.

**Keywords:** COVID-19; Nanozymes; COVID-19diagnosis; Nanozyme-based sensors

## 1. Introduction

On December 31, 2019, the first case of a novel infectious disease with unknown origin (causative agent), features, duration of human transmission, and epidemiological parameters was confirmed in a designated hospital in Wuhan, a major city in China [1, 2]. The studies on this new infectious disease revealed that a new generation of coronavirus, SARS-CoV-2 (severe acute respiratory syndrome coronavirus 2), is its causative agent [3-5]. Coronaviruses are a group of Coronaviridae families with a broad distribution in mammals which are known as the non-segmented positive-sense RNA viruses [6]. This novel disease caused by SARS-CoV-2 was called Coronavirus disease 2019 and termed COVID-19 by WHO on 11 Fed 2020 [7]. Although the human infections resulting from coronavirus are mild in most cases [6], shortly after the first report of COVID-19, the novel COVID-19 exhibited a high potential for outbreaks and becoming an epidemic disease and even a pandemic, as now we see in the world [8-10]. Currently, there are several methods for diagnosis of COVID-19 including real-time reverse transcription-polymerase chain reaction (rRT-PCR), hematology examination, polymerase chain reaction (PCR), diagnostic guidelines based on clinical features, Chest CT scans [1]. Here, it is challenging to develop effective diagnostics and therapeutics against SARS-CoV-2 [11]. Since the first report of COVID-19 in 2019, nanozmyes-based systems have been applied for reliable fast diagnosis of COVID-19. Hence, this mini review focused on overviewing the the nanozyme-based methods toward COVID-19 diagnosis.

## 2. Nanozymes and their biomedical applications

The fast development of nanoscience and material chemistry has increased interest in researching new and innovative synthesis methods to produce new nanomaterials with unique catalytic activity [12, 13], unique optical properties [14-16], high active area [17], antibacterial properties [18], and high biocompatibility [19]. The new field of nanozyme-based catalysis, which has been introduced as an alternative to enzyme-based catalysis, is called nanozyme chemistry. On the other hand, nanozymes are known as nanomaterials with high enzyme-like activity and can be used to simulate enzymatic reactions in harsh environmental conditions (for example, higher temperature or wider pH range) [20-27]. As previously reported in the literature [27], native enzymes, for instance, native peroxidases or ureases suffer from several disadvantages and drawbacks such as low pH stability, low thermal stability, low recoverability, and no reusability. Commonly, to solve these difficulties and drawbacks of native enzymes, the development of enzyme immobilization protocols has been widely considered in the literature [27-30]. Hence to solve these difficulties, the design and development of low-cost nanozymes were considered as an interesting way for performing enzyme-catalyzed reactions in harsh conditions [21, 31]. Nanozymes have been used for several applications in catalysis [32], biomedical imaging [33], tumor therapy [34, 35], and sensing and detection [36-38]. For instance, up to date, different types of nanozyme-based sensors such as single nanozymatic sensors, enzyme-nanozyme hybrid sensors, etc. have been developed [39]. Recently a new generation of nanozyme-based systems called "multinanozyme system' was introduced by Hormozi Jangi et al. (2020) [40, 41]. During the last years, a wide variety of nanozyme-based colorimetric sensors have been developed for the detection and quantification of a variety of analytes for instance, tryptophan [42], glutathione (GSH) [43], dopamine [44], tetracycline [45], metal cations [46], glucose [47], $H_2O_2$ [48], explosives [49], and cysteine [50]. Besides, some of the nanozyme-based senosors with fluorescence-based response had been developed and utilized for detecting several analytes [51, 52].

**3. Nanozymes application for diagnosis of COVID-19**

There is several reports in the literature regarding application of nanozymes for diagnosis of COVID-19. For example, Liang et al. (2021) [53] developed a nanozyme-linked nanosensor for the rapid and quantitative diagnosis of COVID-19 by detecting the SARS-CoV-2 nucleocapsid protein in human blood. Besides, Fu et al. (2021) [54] used porous metallic gold@platinum nanozymes for the diagnosis of COVID-19 via colorimetric detection of spike (S1) protein of SARS-CoV-2, obtaining a wide linear working range over 10–100 ng mL$^{-1}$ along with a low limit of detection (LOD) of 11 ng mL$^{-1}$.

In 2021, Liu et al. [55] developed a paper-based chemiluminescence nanozymatic strip test for sensitive detection of SARS-CoV-2 antigen which the core of their paper test was a Co–Fe@hemin-peroxidase nanozyme that can catalyze chemiluminescence and amplify immune reaction signal. In addition, Liu et al. (2021) [56] introduced a smartphone-based nanozyme-linked immunosorbent assay for quantitative sensing of SARS-CoV-2 nucleocapsid phosphoprotein in 37 serum samples from 20 patients infected with COVID-19. In 2022, Ali and Omer [57] introduced an ultrasensitive aptamer-functionalized Cu-MOF fluorescent nanozyme and utilized it for optical detection of C-reactive protein toward the diagnosis of COVID-19 via colorimetric and fluorometric dual mode responses of a nanozymatic process based on TMB oxidation for colorimetric and variation of fluorescence intensity of Cu-MOF for fluorometric mode detection of COVID-19. Moreover, Zhao et al. (2022) [58] employed MIL-101(CuFe) nanozymes for accurate visual naked-eye diagnosis of COVID-19 via detecting the universal receptor of CD147, providing a very low detection limit of 3 PFU/mL and a detection time as short as 30 min. Besides, Wu et al. (2022) [59] developed a $MnO_2$ nanozyme-mediated CRISPR-Cas12a system for naked-eye diagnosis of COVID-19. In this system, the $MnO_2$ nanorods were initially linked to magnetic beads using a single-stranded DNA (ssDNA). Also, in 2023, He et al. [60] performed a nanozyme-based colorimetric method for naked-eye diagnosis of COVID-19 by iron manganese silicate nanozymes as peroxidase-like nanozymes.

## 5. Conclusions

Several clinical methods had been utilized for diagnosis of COVID-19 for instance, real-time reverse transcription-polymerase chain reaction (rRT-PCR), hematology examination, polymerase chain reaction (PCR), diagnostic guidelines based on clinical features, and Chest CT scans. However, the accurate current methods are time-consuming and expensive and other methods are inaccurate. To solve these drawbacks, nanozyme-based sensors have been developed for the reliable, accurate, and rapid detection of SARS-CoV-2. The main basis of these sensors is the detection of color variation of a nanozyme-mediated oxidation reaction in the presence and the absence of antigens of COVID-19. Besides, some of methods are based on probing the fluorescence of these systems as the clinical signal toward detection of SARS-CoV-2. This mini review focused on overviewing the nanozymes-based methods toward COVID-19 diagnosis. The historical background of COVID-19 was reviewed. Thereafter, the biomedical applications of nanozymes was discussed and finally, the recent progress of early diagnosis of COVID-19 based on nanozymatic systems was briefly reviewed.

**Acknowledgments**

The authors gratefully thank the Hormozi Laboratory of Chemistry and Biochemistry (Zabol, Iran) for the support of this work.

**Conflict of interest**

None.

activity in organic media and enzymatic kinetics of urease and its application for urea removal from water samples. Process Biochemistry, 90, 102-112.

31. Wang, Q., Wei, H., Zhang, Z., Wang, E., & Dong, S. (2018). Nanozyme: An emerging alternative to natural enzyme for biosensing and immunoassay. TrAC Trends in Analytical Chemistry, 105, 218-224.

32. Lu, W., Guo, Y., Zhang, J., Yue, Y., Fan, L., Li, F., ... & Shuang, S. (2022). A High Catalytic Activity Nanozyme Based on Cobalt-Doped Carbon Dots for Biosensor and Anticancer Cell Effect. ACS Applied Materials & Interfaces, 14(51), 57206-57214.

33. Ren, X., Chen, D., Wang, Y., Li, H., Zhang, Y., Chen, H., ... & Huo, M. (2022). Nanozymes-recent development and biomedical applications. Journal of Nanobiotechnology, 20(1), 92.

34. Tang, G., He, J., Liu, J., Yan, X., & Fan, K. (2021, August). Nanozyme for tumor therapy: Surface modification matters. In Exploration (Vol. 1, No. 1, pp. 75-89).

35. Li, Menghuan, Hui Zhang, Yanhua Hou, Xuan Wang, Chencheng Xue, Wei Li, Kaiyong Cai, Yanli Zhao, and Zhong Luo. "State-of-the-art iron-based nanozymes for biocatalytic tumor therapy." Nanoscale Horizons 5, no. 2 (2020): 202-217.

36. Yu, R., Wang, R., Wang, Z., Zhu, Q., & Dai, Z. (2021). Applications of DNA-nanozyme-based sensors. Analyst, 146(4), 1127-1141.
12